\documentclass[twocolumn,showpacs,preprintnumbers,amsmath,amssymb]{revtex4}
\usepackage{graphicx}
\usepackage{dcolumn}
\usepackage{color}
\usepackage{bm}
\begin{document}

\title {``Stochastic Modeling of Coercivity " - A Measure of Non-equilibrium State.}
\author{S. Chakraverty and M. Bandyopadhyay}
\affiliation{Nano Science Unit, S.N. Bose National Center for Basic Sciences,
 JD Block, Sector III, Salt Lake City, Kolkata 700098, India}
\date{\today}

\begin{abstract}
A typical coercivity versus particle size curve for magnetic nanoparticles has been explained by using the Gilbert equation followed by the corresponding Fokker Plank equation. Kramer's treatment has been employed to explain the increase in coercivity in the single domain region. The single to multi-domain transformation has been assumed to explain the decrease in coercive field beyond a certain particle size. The justification for using Langevin theory of paramagnetism (including anisotropy energy) to fit the M vs H curve is discussed. The super-symmetric Hamiltonian approach is used to find out the relaxation time for the spins (making an angle greater than $90^0$ with applied field) at domain wall. The main advantage of our technique is that we can easily take into account the time of measurement as we usually do in realistic measurement.
\end{abstract}
\maketitle
\section{Introduction}
Thermal excitation and relaxation phenomena play a very crucial role in the case of nanoparticles. At finite temperature, single domain nanoparticles often exhibit superparamagnetic behavior, i.e., the relaxation time of the particles is much smaller than the characteristic time scale of the measuring instrument. Superparamagnetic behavior has currently been studied by a number of experimental techniques such as ac and dc susceptibility measurements \cite{dormann,lue,johnsson,dick}, neutron diffraction \cite{mir} etc. These effects have considerable technological interest because of their relevance to the stability of information stored in the form of magnetized particles.\\
Coercivity is an important quantity which plays a crucial role as far as the stabilization of a magnetic system is concerned. Understanding the nature of the coercive field with the variation of particle size is one of the central issues, which has been addressed in the present paper, by using stochastic theory \cite{datta}. 
\begin{figure}[t]
{\rotatebox{0}{\resizebox{5cm}{5cm}{\includegraphics{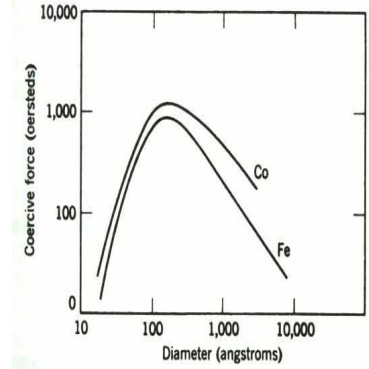}}}}
\caption{\label{fig:exp}Coercive force as a function of particle size.
( W. H. Meiklejohn, Rev. Mod. Phys. 25, 302(1953); F. E. Luborsky, J. Appl. Phys. 32, 171S(1961).)}
\end{figure}
\noindent
The behavior of the coercivity as a function of particle size is a old one\cite{morrish}. The physical phenomena creating the maximum in coercivity of nanoparticles as a function of size is a well known problem in magnetism. This problem has been qualitatively understood. Various theoretical models have been published on the particle size dependence of coercivity \cite{frei,child}. Thermal switching in single-domain particles (where coercivity increases with size of particles) was considered by a lot of authors \cite{brown,aharoni,coffey,klik,garanin}.  These models failed to explain the decrease in H$_c$ with the increase in V at large particle size. Nucleation of domain walls was investigated by Braun \cite{braun1,braun2}. The crossover from single to multi-domain switching was investigated numerically by Hinzke et al \cite{hinzke}. The effect of measurement time (i.e. the time lag between the measurement and the application of field) was not included in their approaches. In this paper we have emphasized to give a mathematical basis to explain this well known phenomena including the effect of measurement time. We quantitatively explain the peak in the coercivity vs. particle size curve. \\
In this paper we have tried to explain the non-monotonic (first increase and then decrease in coercivity, FIG.1) behavior in coercivity against the particle size with the help of non-equilibrium statistical mechanics approach. The time of measurement is automatically included in this approach. Our description is based on the assumption that our system is consist of mono-dispersed particles with no interparticle interaction. Although the particle size distribution and interparticle interaction can produce many interesting effects \cite{sun,sasaki,chakraverty}, but it is beyond the scope of this paper.\\ 
With the preceding background the paper is organized as follows. In Sec. II we have discussed the increasing part of the H$_c$ by assuming the material consist of single domain particle with  high anisotropy barrier limit. This section also contains the magnetization calculation taking into account the anisotropy energy. In Sec.III we have discussed the decreasing part of the coercivity by using super symmetric quantum mechanics (SUSYQM) approach. Finally, in Sec.V we present our main conclusions about the significance of the reported results.\\

\section {Single domain regime}
The subject of how a bulk magnetic specimen acquires a single domain structure and exhibits magnetic viscosity due to Neel relaxation, when its size is reduced, is an old one \cite{neel}. The critical volume of a single domain particle with uniaxial high anisotropy energy as estimated by Kittel \cite{kittel} is $R_c=(9{\sigma}_w/4{\pi}{I_s}^2)$, where $\sigma_w$ is the domain wall energy per unit area and $I_s$ is the saturation magnetization of the particle. The numerical value of $R_c$ comes out to be $\sim 10 \ nm$ for $K\sim 10^ 5$ erg/cc (K is the anisotropy energy per unit volume). Therefore the initial increase in coercivity is observed in single domain particles. This leads us to map our problem to that of an ensemble of uniaxial single domain particles in contact with a heat bath at temperature T. Let us assume that the magnetic moment vector of each particle ($\vec \mu_p$) (FIG. 2) makes an angle $\theta_m$ with the easy axis and this easy axis of the particle is at an angle $\theta_k$ with respect to the applied magnetic  field(${\vec H}_{app}$). The total energy of the system is given by:
\begin{equation}
 E_T=KV{{sin^2(\theta_m)}}-\mu_pH_{app}cos({\theta_k}-{\theta_m}).
\end{equation}
The Gilbert equation \cite{gilbert} governing the dynamics of the spin is
\begin{equation}
\frac{d\vec \mu_p}{dt}={{\gamma_0}\vec \mu_p}\times{\Big[}-\frac{\partial E_T}{\partial {\vec \mu_p}}-\eta
{\frac {d{\vec M}}{dt}}{\Big]}.
\end{equation}

\begin{figure}[t]
{\rotatebox{0}{\resizebox{5cm}{4.5cm}{\includegraphics{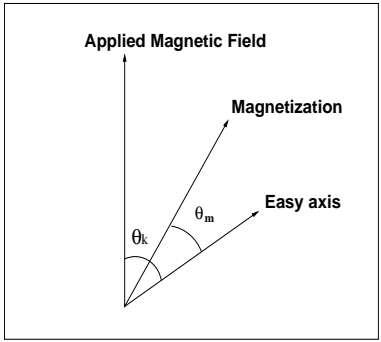}}}}
\caption{\label{fig:structure} Direction of the applied magnetic field, magnetic easy axis and magnetization.}
\end{figure}
\begin{figure}[t]
{\rotatebox{0}{\resizebox{5cm}{5cm}{\includegraphics{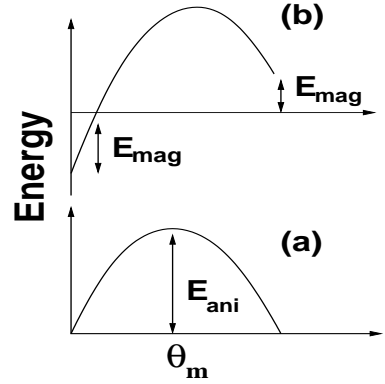}}}}
\caption{\label{fig:structure} Energy profile of an uniaxial single domain magnetic particle (a) in absence and (b) in presence of external magnetic field.}
\end{figure}

The energy profile of the above system (for $H_{app}$=0) is a symmetric double well potential (with two minima at $\theta_1$ and $\theta_2$, generically referred to as $\theta_i$ and a maxima at $\theta_3$(FIG. 3)). Here we shall apply high barrier and weak noise limit to simplify our mathematical calculations. Hence Kramer's ansatz can be used to find the probability of getting a particle making an angle $\theta_m$ with the easy axis and is given by \cite{kramers}
\begin{equation}
P(\theta_m,t) = P(\theta_i,t)e^{-{\frac{({E_T(\theta_m)}-{E_T(\theta_i)})}{k_BT}}}.
\end{equation}
Although we can solve the corresponding Fokker Plank equation with the help of SUSYQM. approach, we have used an approximate method to get a better feeling of the underlying physics of the problem.  
The number of particles oriented between ${\theta_i}-{\theta_m}$ to ${\theta_i}+{\theta_m}$ is given by\\
\begin{equation}
 n_i= P(\theta_i,t) e^{-{\frac {E_T{(\theta_i)}} {k_BT}}} I_i,
\end{equation}
where, 
\begin{equation}
I_i={\int_{{\theta_i}-{\theta_m}}^{{\theta_i}+{\theta_m}}} e^{-{\frac {E_T{(\theta_m)}} {k_BT}}} {sin(\theta_m)}d{\theta_m}, 
\end{equation}
with i = 1 or 2. Since in our case the potential barrier ($\delta E$) is much higher than the 
thermal fluctuation, we have used the saddle point approximation in Eq.(5) which results in
\begin{equation}
I_i=\frac {e^{-{\frac{E_T(\theta_i)}{k_BT}}} k_BT} {C_i},
\end{equation}
where $C_i$ is the curvature at the  i-th minima. 
Now the continuity equation for $P(\theta_m,t)$ is given by
\begin{equation}
\frac {dP(\theta_m,t)}{dt} = - {\vec \nabla . \vec J^{\prime} },
\end{equation}
where,
\begin{equation}
\vec J^{\prime}= {{\vec J}} - k {{\vec\nabla P(\theta_m,t)}}.
\end{equation}
The second term of the above equation represent a diffusion in ${\theta_m}- P({\theta_m}) $ space, which takes care of thermal agitation and $\vec J$=$\vec v P(\theta_m,t)$, where $\vec v = \frac {d{\vec M}}{dt}$.  
We have also assumed the diffusion coefficient to be space independent. In the limit of high relaxation time $J^{\prime}(\theta_m,t)$ becomes independent of $\theta$ and is given by
\begin{equation}
{J_{qs}(t)} = -{\frac {sin(\theta_m)}{\tau_0}}{\Big[}{\frac {1}{k_BT}} {\frac {\partial E_T(\theta_m)}{\partial {\theta_m}}} P(\theta_m,t) + {\frac {\partial {P({\theta_m},t)}} {\partial {\theta_m}}}{\Big]}. 
\end{equation}  
Using Kramer's ansatz in the above equation one can write 
\begin{equation}       
{\frac {d n_2}{dt}} = -{\Big[}{\frac {n_2}{\tau_2}} - {\frac {n_1} {\tau_1}}{\Big]},
\end{equation}
where $\tau_i = \tau_0 I_3 I_i$ and $I_3 = \int_{\theta_1}^{\theta_2} {\frac {e^{\frac {E_T(\theta_m)} {k_BT} }}{sin{\theta_m}}} d{\theta_m}$.
\begin{figure}[t]
{\rotatebox{270}{\resizebox{7cm}{8cm}{\includegraphics{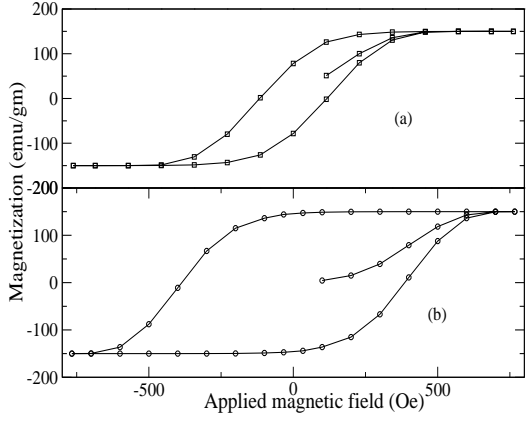}}}}
\caption{\label{fig:structure} Computer simulated magnetization curve of two different particles (a)7 nm (b) 8.5 nm, using Eq. (11).}
\end{figure}
The solution of Eq.(10) is 
\begin{equation}
n_2(t) = \frac {n\tau_2 - e^{-({\frac {1}{\tau_1}}+{\frac {1} {\tau_2}})t} (n\tau_2 - n_2(\tau_2 + \tau_1)) } {\tau_1 + \tau_2},
\end{equation}
where $(n_1-n_2)$ is proportional to the net magnetization along the direction of the applied magnetic field. 
Eq.(11) needs one initial condition i.e. $n_2(0)$ to get the value of $n_2(t)$. For a single domain particle with large relaxation time, if one changes the magnetic field 
after a finite interval of time (t), then
${\lim_{\delta H \to 0^-}}n_{2}^{H-\delta H}(t) = n^{H^\rightarrow}_{2}(0)\neq {\lim_{\delta H \to 0^+}}n_{2}^{H+\delta H}(t) =  n^{H^\leftarrow}_{2}(0) $. This implies that for a particular value of H one should not expect to get the same value of magnetization during increasing and decreasing cycle of H. Since the relaxation time $\tau_i$ increases with particle volume, $M^{H\leftarrow} - M^{H\rightarrow}$ also increases with particle volume giving rise to higher coercivity. Hence coercivity is a  consequence of the quasi-static non-equilibrium measurement. Therefore, Langevin theory of paramagnetism is not applicable in these cases. We have used Eq.(11) to numerically generate M vs H curve as shown in FIG.4, for particle size 7 nm  and 8.5 nm. We have used t = 50sec and K=10$^5$erg/cc in the numerical calculation, which is realistic for measurements of coercivity by vibrating sample magnetometer.\\
 In case of a superparamagnetic system, a common practice is to fit the magnetization curve by using Langevin theory of paramagnetism \cite{kodama}. 
But it is not quite obvious, since Langevin theory of paramagnetism does not include the magnetocrystaline anisotropy energy. Therefore an attempt has been made here to calculate the magnetization of a superparamagnetic sample  at equilibriums. Magnetization\\
\begin{equation} 
<M>=\mu_p<cos\theta>, 
\end{equation}
where,
$<cos\theta>={{\int_0}^{\pi}} P(\theta,H_{app})\ cos\theta \ sin\theta \ d\theta $, \\
 
$P(\theta,H_{app}) = \frac{e^{\frac{-E_T(\theta,H)}{k_BT}}}{Z}$,\\ 

$E_T= - H \mu_p \ cos\theta + k \ sin^2{\theta}$, \\
and partition function

\begin{eqnarray}
Z & = & \sqrt{\frac{\pi}{k_1}}{\frac{e^{-(k_1+{\frac{\alpha^2}{4k_1}})}}{2}}(Erfi[ \sqrt k_1({\frac{\alpha}{2k}}+1)]\nonumber \\
&{}&-Erfi[\sqrt k_1({\frac{\alpha}{2k}}-1)]) \nonumber ,\\
\end{eqnarray}
where $k=KV$ (V is the volume of the particle), $\alpha = \frac {H_{app}\mu_p}{k_BT}$ and $k_1=\frac{k}{k_BT}$. Here Erfi is the error function \cite{grad}. The magnetization of the system 
\begin{equation}
<M>=\mu_p S[k_1,\alpha]-{\frac {\alpha}{2k_1}},
\end{equation}
 with
\begin{equation}
S[k_1,\alpha]  = \mu_p \frac {e^{k_1{(\frac {\alpha}{2k_1}+1})^2}-e^{k_1{(\frac {\alpha}{2k_1}-1})^2}} {X}, 
\end{equation}
where $ X= {\sqrt{k_1\pi}(Erfi[\sqrt k_1({\frac{\alpha}{2k}}+1)]-Erfi[\sqrt k_1({\frac{\alpha}{2k}}-1)])}.$ 
Now $<M>$ $\rightarrow$ $\mu_{p} L(\alpha)$ at the limit $ k_1 $ $\rightarrow$ 0, where $L(\alpha)$ is the Langevin function. Therefore in general it is better to use Eq.(14) rather than Langevin theory of paramagnetism. But for very small particle size one can use Langevin theory of paramagnetism directly.\\
\noindent
One can calculate the magnetization for the superparamagnetic state using the rate equation at equilibrium for the particles whose barrier height is at least 5 times larger than the thermal energy. Let us consider the situation where we have applied a dc magnetic field parallel ($\theta_k$=0) to the easy axis. If $n_2$ and $n_1$ be the number of particles having spins parallel ($\theta_k=0$) and antiparallel ($\theta_k=180^0$) to the applied field, the magnetization along the direction of magnetic field is $M_s(n_2-n_1)$, $M_s$ being the saturation magnetization. From the master equation at equilibriums we have
\begin{equation}
<M(n_2-n_1)>= M_s (\frac {{\tau_2} - \tau_1}{\tau_1})\frac {e^{-{\frac {E_1}{k_BT}}}} {e^{-{\frac {E_1}{k_BT}}} + e^{-{\frac {E_2}{k_BT}}}},
\end{equation}
where
$E_1=E_{mag}=-E_2$ and $\tau_i=\tau_0 e^{\frac {E_{ani}-E_i}{k_BT}}$. The above equation gives us 
\begin{equation}
<M_s>= M_s \frac{e^{\frac{E_{mag}}{k_BT}}-e^{-{\frac{E_{mag}}{k_BT}}}}{e^{\frac{E_{mag}}{k_BT}}+e^{-{\frac{E_{mag}}{k_BT}}}}.
\end{equation}
This indicates that the anisotropy energy does not affect the magnetization of superparamagnetic particle (with large anisotropy energy) in equilibrium. 
\section {Multi-domain Regime - SuperSymetric Quantum Mechanics (SUSYQM) Approach}
Let us now concentrate on the second region where the coercivity field decreases with the increase in particle size. It is clear from the above discussion that this can not happen if the particles still comprise single domains. Since the coercivity of the single domain particle increases monotonically with the increase in particle volume,hence a single to multi domain transformation takes place at the maximum of coercivity.\\
To explain the decreasing part of the coercivity field let us consider an arrangement of spin in a linear chain as shown in FIG.5. In the following discussion one should keep in mind  that we are not interested here in the origin of the domain wall, but we assume the existence of domain and the Hamiltonian contains the relevant terms. The Gilbert equation corresponding to i-th spin is 
\begin{equation}
\frac{d\vec \mu^i}{dt}={{\gamma_0}\vec \mu^i}\times{\Big[}-\frac{\partial E_T}{\partial {\vec \mu^i}}-\eta
{\frac {d{\vec \mu^i}}{dt}}{\Big]}.
\end{equation}
Let us consider cylindrical polar coordinates, where $z = 0$ represents 0th site, $z = a$ represents 1st site and so on. For a particular site the spin can move over the surface of the cylinder along a semicircular curve $(\theta \epsilon [0,\pi])$.  
\begin{figure}[t]
{\rotatebox{0}{\resizebox{5cm}{4cm}{\includegraphics{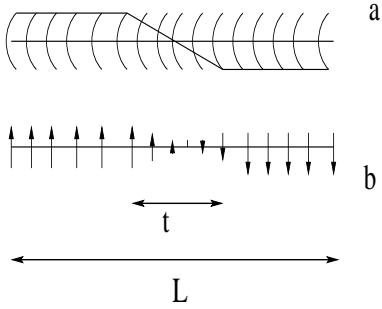}}}}
\caption{\label{fig:structure} Spin distribution near $180^0$ domain wall (a) at real space (b) at probability space.}
\end{figure}
Writing everything in terms of cylindrical polar coordinate and proceeding as above we have
\begin{equation}
\frac{d {P^i}(\theta,t)}{dt} = -h' k_BT{\frac{\partial}{\partial \theta}}{\Big[}{\frac{1}{k_BT}}{\frac{\partial V(\theta)}{\partial \theta}}{P^i}(\theta, t) - {\frac {\partial {P^i}(\theta, t)}{\partial \theta}}{\Big]},
\end{equation}
with $h'=-\frac {{\gamma_0}^2 \eta {\mu}^i} {1 + {\gamma_0}^2 \eta {\mu^i}^2}$,
subjected to the condition $\frac {J^i(\theta,t)}{dz}=0$, which physically means that there is no spin hopping between two sites. This implies that the spin will start relaxing along the surface of the cylinder without changing its position along Z-axis. To solve the above equation let us perform the following transformation $P(\theta,t)= \sqrt{P_{eq}(\theta)}{\psi({\theta,t})}$. Where $P_{eq}(\theta)=A_0e^{\frac{-V{(\theta)}}{\epsilon}}$ and $\epsilon = k_BT$. Using the transformation in Eq.(19) we have $\frac{d\psi}{dt}= k_1{\psi}''+h_1{\Big(}{\frac{V''(\theta)}{2}}-{\frac{{V'}^2}{4\epsilon}}{\Big)}\psi$. Introducing a new function $\phi(\theta)$ such that $\psi(\theta,t) = \phi(\theta)e^{-{\lambda}' t}$ , we obtain
\begin{equation}
\lambda \phi = \epsilon {\phi}'' + {\Big(}{\frac{V''(\theta)}{2}}-{\frac{{V'}^2}{4\epsilon}}{\Big)} \phi,
\end {equation}
with $\lambda=\frac{{\lambda}'}{h_1}$ defining two operator $A=\frac{\partial}{\partial \theta}+\frac{V'}{2\epsilon}$ and $A^{\dagger}=-\frac{\partial}{\partial \theta}+\frac{V'}{2\epsilon}$, such that $\epsilon A^{\dagger}A \phi = \lambda \phi$ with ground state eigenvalue equal to zero (since $A\phi_0=0$ to get equilibrium distribution). It can be shown that if $\phi_1$ be the first excited eigenstate of $A^{\dagger}A$ then it is the ground state of $AA^{\dagger}$ with ground state eigenvalue ${\lambda_1}$ \cite{cooper}. Now one can apply variational method to get $\lambda_1$
\begin{equation}
\lambda_1=\frac{\int \phi_1({\theta})\epsilon AA^{\dagger}\phi_1(\theta)d\theta}{\int \phi_1({\theta})\phi_1({\theta})d\theta}.
\end{equation}
Now we are in a position to get the solution of the Fokker Planck equation. Suppose we have applied a static magnetic field along the crystallographic direction of the sample. The potential energy of the spins making an angle less than $90^0$ can be approximated by harmonic oscillator like potential $\frac {1}{2} K\theta^2$. Which gives us the relaxation time of the order of $h_1(\epsilon - K)$. The spins at the end of the domain wall have a double well like potential. For such double well potential we have $\lambda_1\sim h_1(e^{-(V_0-V(\theta_1))}+ e^{-(V_0-V(\theta_2)}),$ where $V_0$ is the barrier height and $\theta_1$ and $\theta_2$ are the position of the two minima. Now it is clear that the relaxation time depends on the damping parameter as well as the barrier height, which in turn depends on the value of anisotropy constant and the angle between successive spins. The anisotropy constant is higher (one order of magnitude) for smaller particle. So for a smaller particle it takes more time to relax back to its initial configuration, giving rise to a higher coercive field. The above model indicates that our system under consideration, consist of a linear chain of ferromagnetic particles having two domains, with their easy axes parallel to each other and with applied magnetic field also. The above model also does not contain the domain of closure. Still the model can be regarded as the starting point to explain qualitatively the hysteresis of a multi domain system. \\ 
\section{conclusion}
The particle size dependence of the coercive field is explained from the view point of non-equilibrium statistical mechanics. The zero hysteresis loss  indicates that the system reaches its equilibrium state before the next data is taken. It is shown that the increase in coercivity is the effect of increase in relaxation time. The numerical calculation of hysteresis curve shows that the loss as well as the coercive field increases with particle size. During the numerical calculation assumption has been made that $k_BT< \delta E$. The Langevin theory of paramagnetism for superparamagnetic particle has been reestablished, after taking into account the magnetocrystaline anisotropy effect. Assuming the single to multi-domain transformation we have shown that the relaxation time of the samples decreases with increases in particle size due to decrease in surface pressure and anisotropy constant which gives a decrease in coercivity. The important point to note that in case of any experimental study of a single domain particle, one should perform this coercive field ($H_{c}$) vs. particle size curve and figure out the peak in the curve. One should perform all the measurements below this peak particle size to analyze the behavior of single domain magnetic nanoparticles. 

\section*{Acknowledgments}   
We wishes to express  gratitude to Professor Sushanta Dattagupta for his illuminating course in non-equilibriums statistical mechanics and to Professor Binayak Dutta Ray for helping us to understand the underlying physics. We also grateful to Professor H. S. Mani for the constant encouragement. Financial support from the Department of Science and Technology, Government of India is gratefully acknowledged. \\

\end{document}